\documentclass[draftclsnofoot,onecolumn]{IEEEtran}

\usepackage{lineno,hyperref}
\modulolinenumbers[5]
\usepackage{geometry}
\usepackage{hyperref}
\usepackage{amsfonts}
\usepackage{amsmath}
\usepackage{amsthm}
\usepackage{algorithm}  
\usepackage{algpseudocode}  
\usepackage{amsmath}  
\usepackage{graphicx}
\usepackage{subfigure}
\usepackage{booktabs}  
\usepackage{multicol}  
\usepackage{multirow}
\usepackage{threeparttable}  
\usepackage{makecell}
\usepackage{caption2}
\usepackage{lipsum}

\date{} 

\newtheorem{theorem}{Theorem}

\newtheorem{definition}{Definition}

\ifCLASSINFOpdf

\else

\fi

\begin{document}

\title{Tensor Completion via Leverage Sampling and Tensor QR Decomposition for Network Latency Estimation \textsc{}}

\author{\IEEEauthorblockN{Jun Lei\IEEEauthorrefmark{1},
		Ji-Qian Zhao\IEEEauthorrefmark{1},
		Jing-Qi Wang\IEEEauthorrefmark{1}, 
		An-Bao Xu \IEEEauthorrefmark{2}{$ ^{\#} $} \thanks{ $ ^{\#} $ Corresponding author}}

\IEEEauthorblockA{\IEEEauthorrefmark{1}College of Computer Science and Aritificial Intelligence, Wenzhou University, Zhejiang 325035, China.}
\IEEEauthorblockA{\IEEEauthorrefmark{2}College of Mathematics and Physics, Wenzhou University, Zhejiang 325035, China.}

\thanks{This work was supported by the National Natural Science Foundation of
	China under Grant 11801418.

E-mail address: leik5102@gmail.com (Jun Lei), xuanbao@wzu.edu.cn (An-Bao Xu).
}

}


\markboth{  }%
{Shell \MakeLowercase{\textit{et al.}}: Bare Demo of IEEEtran.cls for Journals}
\maketitle

\begin{abstract}
In this paper, we consider the network latency estimation, which has been an important metric for network performance. However, a large scale of network latency estimation requires a lot of computing time. Therefore, we propose a new method that is much faster and maintains high accuracy. The data structure of network nodes can form a matrix, and the tensor model can be formed by introducing the time dimension. Thus, the entire problem can be be summarized as a tensor completion problem. The main idea of our method is improving the tensor leverage sampling strategy and introduce tensor QR decomposition into tensor completion. To achieve faster tensor leverage sampling, we replace tensor singular decomposition (t-SVD) with tensor CSVD-QR to appoximate t-SVD. To achieve faster completion for incomplete tensor, we use the tensor $L_{2,1}$-norm rather than traditional tensor nuclear norm. Furthermore, we introduce tensor QR decomposition into alternating direction method of multipliers (ADMM) framework. Numerical experiments witness that our method is faster than state-of-art algorithms with satisfactory accuracy.
\end{abstract}



\IEEEpeerreviewmaketitle

\section{Introduction}

\IEEEPARstart{N}{etwork} latency estimation has played an important role in network performance evaluation and has been widely concerned in recent years. It is known to all that many applications are sensitive to latency \cite{HP2019}. For instance, to ensure the quality of service (QoS), the low transmission latency is extrmely required by mobile video calls. Thus, it is crucial to acquire the latencies of the whole network within a short time duration \cite{SM2014}. However, It is not practical to measure the whole large-scale network latency comprehensively due to the extremely high cost, as a result of which, we need a fast and precise method of network latency estimation.

Matrix completion (MC) \cite{ZB2017,GM2012,IK2018,XD2020,XA2017} and tensor completion (TC) \cite{LR1993,TR1966,GM2015,WV2017,XA2020} are the main shunt of network latency estimation in previous work. MC was first proposed by Candè and Recht \cite{CF2008}, and was applied to network latency estimation \cite{XL2015,ZH2015} because of the low-rank structure of network traffic data \cite{MY2004}. TC is an extension of MC with higher precision and higher complexity \cite{DZ2020}. The purpose of TC is to complete the missing entries from partially observed subset. Hence, TC introduces the dimension of time, but the time will not be too long in latency estimation. In the field of signal processing and data analysis, TC has been an advanced technology widely used. For example, TC has been applied to network traffic estimation \cite{XL2018,ZM2009,MA2010}. 

In the past few years, many scholars have tried to optimize network latency estimation. In terms of MC, Zhu \cite{ZB2017} proposed a latency recovery scheme by using a subset of source-destination pairs. Xie \cite{XL2015} proposed an adaptive sampling scheme to improve the accuracy of it. Mao and Saul \cite{MY2004} predicted the latencies of large-scale networks by matrix factorization. Liao \cite{LW2012} put forward a new algorithm called decentralized matrix factorization by stochastic gradient descent (DMFSGD). In terms of TC, Zhou \cite{ZH2015} proposed expliot the advantage of tensor CP decomposition for network traffic estimation. Liu \cite{LD2015} put forward a TC strategy through "flattening" the tensor. Deng \cite{DZ2020} artfully used the tensor completion model based on t-SVD \cite{KD2011} and an effective sampling strategy that greatly improve the precision of network latency. Lin \cite{LD2022} proposed a new robust spatial-temporal graph-tensor recovery model with high stability. There are also works with respect to approximating the underlying low-tubal-rank tensor by adaptive tubal sampling strategies \cite{ZM2019,LS2015} and uniform element-wise sampling strategies \cite{LS2016} that make great sense. These approaches including other excellent related methods have made an important contributions to this field.

Most of the researchers in this field pay more attention to the accuracy of network delay estimation but do not pay much attention to the speed of the algorithm to some extent. It is meaningful to enhance the precision but network latency estimation also needs a strong real-time performance. Different from the above schemes, the purpose of our method is to greatly improve the speed while maintaining high accuracy. Inspired by \cite{DZ2020}, we improve the tensor leverage sampling strategy and tensor comopletion method, and we exploit tensor QR decomposition as a core subroutine to approximate tensor singular value decomposiion (t-SVD), which is the core subroutine in ADMM framework \cite{LJ2018}. Furthermore, we use tensor $L_{2,1}$-norm \cite{ZA2020} to solve the tensor completion optimization problem with a fast speed and high accuracy.

The main contributions of this paper are:
\begin{itemize}
	\item We propose a new method named QRLS, which is an improvement of tensor leverage sampling with faster speed. It replace the t-SVD in tensor leverage sampling with iterative QR decomposition to approximate t-SVD.
	
	\item We introduce tensor $L_{2,1}$-norm into network tensor completion model. Tensor factorization is an frequent process to compute the nuclear norm with a huge time cost while tensor $L_{2,1}$-norm can improve the speed of this process. In addition, the accuracy of tensor $L_{2,1}$-norm is also satisfactory because of its low computational complexity and high robustness to noise compared to nuclear norm.
	
	\item We propose a new model called LNLS-TQR that is a combination of tensor completion based on $L_{2,1}$-norm and a fast leverage sampling method. The QR decomposition is the core subroutine of the above methods. Our method can rapidly and precisely obtain informative node pairs and can accurately estimate numerous unobserved node pairs. Numerical experiments witness that our method is not only faster than some of excellent algorithms proposed recent years, but it also has satisfactory accuracy.
\end{itemize}

\section{\large\bf Notations and preliminaries}
\subsection{Notations}
In this papper, handwritten uppercase letters represent third-order tensors, printed uppercase letters represent matrices,and printed lowercase letters represent vectors. For example, ${\cal X} \in {\mathbb{R}^{{n_1} \times {n_2} \times {n_3}}}$ represents a third order tensor, $X \in{\mathbb{R}^{{n_1} \times {n_2}}}$ is a matrix and $x \in {\mathbb{R}^{{n_1}}}$ is a vector. The definition method of matlab is adopted for the lower corner label. For instance, ${\cal X}(k,:,:)$, ${\cal X}(:,k,:)$, ${\cal X}(:,:,k)$ means horizontal, lateral and frontal slice of a third-order tensor respectively. ${\cal X}(i,j,:)$ denotes the tube of ${i^{th}}$ row, ${j^{th}}$ column when faced with frontal slices of ${\cal X}$ and ${\mathcal{X}_{ijk}}$ denotes the ${(i,j,k)^{th}}$ entry of a tensor. If the whole entries of a tensor is zero, the tensor will be denoted by $\mathcal{O}$. A tensor with its first slice being an identity matrix and others being zero matrices is defined as an identity tensor $\mathcal{I}$. Specially, ${{\cal X}^{(k)}}$ denotes the ${k^{th}}$ frontal silice of ${\cal X}$. The Frobenius norm of a tensor ${\mathcal X}$ is $\parallel \mathcal{X}{\parallel _F}{\rm{ = }}\sqrt {\sum\nolimits_{i,j,k} {{{\left| {{\mathcal{X}_{ijk}}} \right|}^2}} } $.

\subsection{Background knowledge}

T-product is the product of tensors. If a tensor ${\cal X}$ is the t-product of ${\cal A}$ and ${\cal B}$, it will be denoted as ${\cal X} = {\cal A}*{\cal B}$. Before we understand the t-product, we need some background knowledge.

\begin{definition}\label{Discete Fourier transformation}
	\textbf{(Discrete Fourier Transformation (DFT))}
	Discrete Fourier transformation is a linear transformation on a vector that satifies the mapping \[\mathbf{fft}:{\mathbb{C}^n} \to {\mathbb{C}^n}\]\[v \to \hat v\] where $v \in {\mathbb{C}^n}$ and the matrix for the transformation is:
	\begin{center}
		${F} _{n} = \begin{bmatrix}
			1& 1 & 1& \cdots &1\\
			1& \omega  & \omega ^{2}  & \cdots &\omega ^{n-1} \\
			\vdots & \vdots  & \vdots  & \ddots &\vdots \\
			1& \omega ^{n-1} & \omega ^{2\left ( n-1 \right )} &\cdots&\omega ^{\left ( n-1 \right )\left ( n-1 \right )}
		\end{bmatrix} \in \mathbb{C} ^{n\times n} $,
	\end{center}
	where $\omega$ is a complex number that is equal to $e^{-\frac{2\pi i}{n} }$ and ${F} _{n}$ will operate on a n-dimensional column vectors. 
\end{definition}

From Definition \ref{Discete Fourier transformation}, we can infer that $\frac{{{F_n}}}{{\sqrt n }}$ is unitary, i.e.,
\begin{equation}\label{2.1}
	F_{n}^{\ast}F_{n}=F_{n}F_{n}^{\ast}=nI_{n},
\end{equation}
where $F_n^ * $ is the conjugate transpose of $F_{n}$. We can apply discrete Fourier transformation to a vector in the third dimension (tube) of the tensor ${\cal X}$  and express it as $\hat {\cal X} = \mathbf{fft}({\cal X},3)$ where 3 in the bracket means the third dimension. Of course, we have ${\cal X} = \mathbf{ifft}(\hat {\cal X},3)$ where $\mathbf{ifft}$ is the inverse of $\mathbf{ifft}$.

Assuming that there's a matrix $\tilde X \in {\mathbb{C}^{^{{n_1}{n_3} \times {n_2}{n_3}}}}$ satisfying the following form:
\begin{center}
	$\hat X = \mathbf{bdiag}\left({\hat {\cal X}}\right) = \begin{bmatrix}
		\mathcal{\hat X}^{\left ( 1 \right ) }&  &   & \\
		& \mathcal{\hat X}^{\left ( 2 \right ) } &  & \\
		&  & \ddots  & \\
		&  &  &\mathcal{\hat X}^{\left ( n_{3} \right ) }
	\end{bmatrix}$,
\end{center}
where $\mathbf{bdiag}$ is an operator that can transform a tensor into a block diagonal matrix like above. Furthermore, the block circulant matrix is defined by \begin{center}
	$\mathbf{bcirc}\left (\mathcal{X}\right ) = \begin{bmatrix}
		\mathcal{X}^{\left ( 1 \right )}   &  \mathcal{X}^{\left ( n_{3} \right )} & \cdots  &  \mathcal{X}^{\left ( 2 \right )}\\
		\mathcal{X}^{\left ( 2 \right )} &  \mathcal{X}^{\left ( 1 \right )} & \cdots  &  \mathcal{X}^{\left ( 3 \right )}\\
		\vdots  & \ddots  & \ddots & \vdots\\
		\mathcal{X}^{\left ( n_{3} \right )} &  \mathcal{X}^{\left ( n_{3}-1 \right )} & \cdots & \mathcal{X}^{\left ( 1 \right )}
	\end{bmatrix}$	.
\end{center}

We can exploit DFT to block diagonalize the block circulant matrix and then we have
\begin{equation}\label{2.2}
	\left ( F _{{n}_{3}}\otimes I_{{n}_{1}}  \right ) \cdot \mathbf{bcirc}\left ( \mathcal{X}  \right )\cdot \left ( F _{{n}_{3}}^{-1}\otimes I_{{n}_{2}} \right )  =\tilde{X}   ,
\end{equation}
where $\otimes$ is the Kronecker product.
\begin{definition}\label{t-product}\cite{KD2011}
	\textbf{(T-product )} Assuming that $\mathcal{X}\in \mathbb{C}^{n_{1}\times n_{2}\times n_{3}}$ and $\mathcal{Y}\in \mathbb{C}^{n_{2}\times l\times n_{3}}$, we can operate t-product $\mathcal{X}\ast\mathcal{Y}$ and get a new tensor with the size of $n_{1}\times l\times n_{3}$:
	\begin{equation}
		\mathcal{X}\ast\mathcal{Y}=\mathbf{fold}\left(\mathbf{bcirc}\left(\mathcal{X}\right)\cdot\mathbf{unfold}\left(\mathcal{Y}\right)\right),
	\end{equation}
	where $\mathbf{unfold}\left( \cdot \right)$ is the operator that arranges the tensor in frontal slices into a long matrix and $\mathbf{fold}\left( \cdot \right)$ is to restore the long matrix to its original tensor, i.e.,
	\begin{center}
		$\mathbf{unfold}\left(\mathcal{X}\right)=\begin{bmatrix}
			\mathcal{X}^{\left ( 1 \right ) }  \\
			\mathcal{X}^{\left ( 2 \right ) } \\
			\vdots  \\
			\mathcal{X}^{\left ( n_{3} \right ) } 
		\end{bmatrix}\in \mathbb{C} ^{n_{1}n_{3}\times n_{2}} ,\quad \mathbf{fold}\left(\mathbf{unfold}\left(\mathcal{X}\right)\right)=\mathcal{X}$.
	\end{center}
\end{definition}

We can see that t-product is similar to the matrix product, except that multiplying the elements of the matrix operation is extended to a circular convolution between tubes in a tensor. According to the details in \cite{KD2011}, we can infer that t-product can be thought of as a matrix product in the Fourier domain.

\begin{definition}\cite{KD2011}
	\textbf{(Tensor conjugate transpose )} Assuming that there's a tensor $\mathcal{X}\in \mathbb{C}^{n_{1}\times n_{2}\times n_{3}}$, the conjugate transpose of $\mathcal{X}\in \mathbb{C}^{n_{1}\times n_{2}\times n_{3}}$ is given by conjugate transposing every frontal slice of $\mathcal{X}\in \mathbb{C}^{n_{1}\times n_{2}\times n_{3}}$ and reversing the order of these slices from the second slice through ${n_3}$. The conjugate transpose of $\mathcal{X}\in \mathbb{C}^{n_{1}\times n_{2}\times n_{3}}$ is denoted by $\mathcal{X}^{\ast}\in \mathbb{C}^{n_{2}\times n_{1}\times n_{3}}$, i.e.,
	\begin{equation}\label{2.3}
		{({\cal X}(:,:,k))^*} = {{\cal X}^*}(:,:,{n_3} - k + 2),\quad k = 2,3, \cdots ,n_{3}
	\end{equation}
\end{definition}

Besides, there are lots of other tensor's operations that is similar to matrix operations \cite{KD2011}. For instance, an orthogonal tensor $\mathcal{X}\in \mathbb{R}^{n_{1}\times n_{2}\times n_{3}}$ satisfies ${\cal X}*{{\cal X}^*} = {{\cal X}^*}*{\cal X} = {\cal I}$, where ${\cal I}$ is the identity tensor. In addition, a tensor with every frontal slice being a diagonal matrix is called a f-diagonal tensor.

\begin{figure}[H]
	\centering
	\noindent\makebox[\textwidth][c] {
		\includegraphics[scale=0.4]{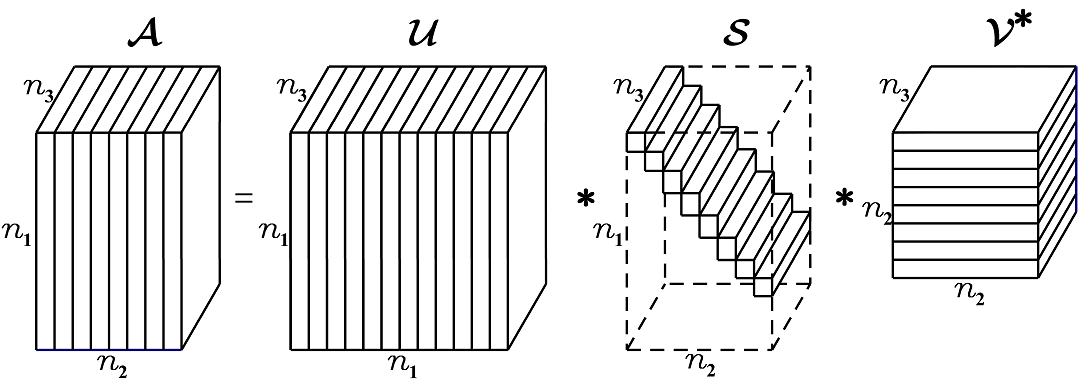} }
	\caption{The illustration of t-SVD of a $n_{1}\times n_{2}\times n_{3}$ tensor.}
	\label{t-SVD}
\end{figure}

\begin{theorem}\label{T-SVD}\cite{KD2011}
	\textbf{(T-SVD  and tensor tubal-rank )} Assuming that there is a tensor $\mathcal{X}\in \mathbb{R}^{n_{1}\times n_{2}\times n_{3}}$ and then we can decompose $\mathcal{X}$, i.e.,
	\begin{equation}
		\mathcal{X}=\mathcal{U}\ast\mathcal{S}\ast\mathcal{V}^{\ast},
	\end{equation}
	where both $\mathcal{U}\in \mathbb{R}^{n_{1}\times n_{1}\times n_{3}}$ and $\mathcal{V}\in \mathbb{R}^{n_{2}\times n_{2}\times n_{3}}$ are orthogonal tensors, and $\mathcal{S}\in \mathbb{R}^{n_{1}\times n_{2}\times n_{3}}$ is a f-diagonal tensor, as shown in Figure \ref{t-SVD}. The complexity of t-SVD is $\mathit{O}\left(20n_{1}n_{2}^{2}n_{3} \right)$. The tubal-rank of ${\cal X}$ is the number of nonzero tubes of ${\cal S}$.
\end{theorem}

\begin{figure}[H]
	\centering
	\noindent\makebox[\textwidth][c] {
		\includegraphics[scale=0.4]{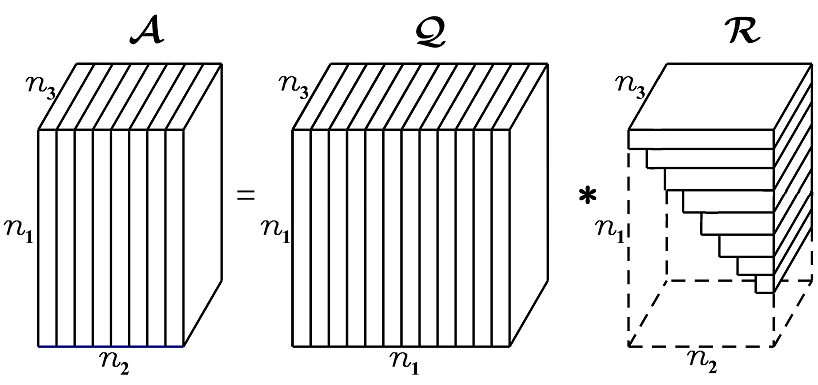} }
	\caption{The illustration of t-QR of a $n_{1}\times n_{2}\times n_{3}$ tensor.}
	\label{t-QR}
\end{figure}

\begin{theorem}\label{T-QR}\cite{KM2013}
	\textbf{(T-QR )} Assuming that there is a tensor $\mathcal{X}\in \mathbb{R}^{n_{1}\times n_{2}\times n_{3}}$, and then $\mathcal{X}$ can be decomposed, i.e.,
	\begin{equation}
		\mathcal{X}=\mathcal{Q}\ast\mathcal{R}
	\end{equation}
	where $\mathcal{Q}\in \mathbb{R}^{n_{1}\times n_{1}\times n_{3}}$ is an orthogonal tensor, and $\mathcal{R}\in \mathbb{R}^{n_{1}\times n_{2}\times n_{3}}$ is similar to the upper triangular matrix, as shown in Figure \ref{t-QR}. The complexity of t-QR is $\mathit{O}\left(2n_{1}n_{2}^{2}n_{3} \right)$.
\end{theorem}

In recent years, t-SVD is one of the most popular strategies of tensor factorization widely applied in the tensor completion problem because of its high accuracy with pretty fast speed. Compared with Theorem \ref{T-SVD} and Theorem \ref{T-QR}, it's easy to find that t-QR is ten times faster than t-SVD. If we can insert t-QR into the tensor completion problem, the algorithm will be more cost-efficient.

\begin{theorem}\label{T-Leverage score}\cite{DZ2020}
	\textbf{(Tensor leverage scores )} Let $\mathcal{X}\in \mathbb{R}^{n_{1}\times n_{2}\times n_{3}}$ with tubal-rank $r$. The leverage scores ${u_i}({\cal X})$ for the ${i^{th}}$ horizontal slice ${\cal X}(i,:,:)$ for the ${j^{th}}$ lateral slice ${\cal X}(:,j,:)$ can be expressed as:
	\begin{equation}
		\begin{array}{l}
			{u_i}({\cal X}) = \frac{{{n_1}}}{r}\left\| {{{\cal U}^*}*{{\vec e}_i}} \right\|_F^2,i = 1,2, \cdots ,{n_1}\\
			{v_j}({\cal X}) = \frac{{{n_2}}}{r}\left\| {{{\cal V}^*}*{{\vec e}_j}} \right\|_F^2,j = 1,2, \cdots ,{n_2}
		\end{array}
	\end{equation}
	where ${{\vec e}_i} \in {\mathbb{R}^{{n_2} \times 1 \times {n_3}}}$ satisfies the entry ${{\vec e}_{i11}} = 1$ and other entries are zeros. Namely, ${u_i}({\cal X})$ is Frobenius norm of ${i^{th}}$ lateral slices of ${\cal U}$ and ${v_j}({\cal X})$ is Frobenius norm of ${j^{th}}$ lateral slices of ${\cal V}$ where $\mathcal{U}$ and $\mathcal{V}$ comes from t-SVD in Theorem \ref{T-SVD}.   
\end{theorem}

The purpose of leverage sampling strategy is to collect informative sample points under the circumstance of limited sampling cost and the leverage score is an indication of the amount of information. In the matrix leverage sampling, it is thought that the singular vector contains a lot of information in the original matrix and quantize it by Frobenius norm of singular vector. Hence, we can generalize this property to tensors \cite{DZ2020}, as a result of which, tensor leverage sampling strategy becomes an excellent algorithm with high precision in a high dimension.

\subsection{Problem description}

The device-to-device network latency information can be stored in a matrix and this matrix is not necessarily symmetric. In addition, to obtian the nodes with more information, we need to introduce the time dimension into the model \cite{DZ2020}, and thus the network latency estimation problem can be reduced to the third-order tensor completion problem. For example, the entry ${\cal X}(i,j,k)$ denotes the network latency from the ${i^{th}}$ device to the ${j^{th}}$ one in the ${k^{th}}$ time slot, but we cannot measure all the latency information in a large scale. The remaining unmeasured information that needs to be estimated is the tensor element that needs to be completed.

The tensor completion problem can be described as follow:
\begin{equation}\label{1.1}
	\min_{\mathcal{X}\in \mathbb{R}^{n_{1}\times n_{2}\times n_{3} }   } \text{rank}_{\text{t}} \left ( \mathcal{X}  \right )   \quad \text{s.t.} \quad \mathcal{X}_{i,j,k}= \mathcal{M}_{i,j,k}, \quad \left ( i,j,k \right ) \in  \Omega  
\end{equation}
where $ \mathcal{X} \in \mathbb{R}^{n_{1}\times n_{2} \times n_{3}}$ is a low-tubal-rank tensor,  $ \text{rank}_{ \text{t}} \left ( \mathcal{X}  \right )$ is the tubal-rank of $ \mathcal{X}$, $ \mathcal{M} \in \mathbb{R}^{n_{1} \times n_{2} \times n_{3}}$ the tensor that we need to complete, and the set $ \Omega$ stores the tensor index we have known. In addition, we define a tensor $\mathcal{A}\in \mathbb{R}^{n_{1}\times n_{2}\times n_{3} }$ as follow :
\begin{equation}\label{Aijk}
	\mathcal{A}_{ijk}{\rm{ = }}\left\{ \begin{array}{l}
		1,\quad \left( {i,j,k} \right) \in \Omega \\
		0,\quad else.
	\end{array} \right.
\end{equation}
From (\ref{1.1}) and (\ref{Aijk}), it is easy to realize that 
\begin{equation}
	\mathcal{M} = \mathcal{X} \circ \mathcal{A},
\end{equation}
where $\circ$ denotes Hadamard product.

Since Problem \ref{1.1} is NP hard and the objective function ${f} \left ( \mathcal{X}  \right ) = \text{rank}_{\text{t}} \left ( \mathcal{X}  \right )$ is not convex, we usually choose a surrogate convex function \cite{KM2013} becoming the objective function, and then Problem (\ref{1.1}) can be transformed as follow:
\begin{equation}\label{1.2}
	\min_{\mathcal{X}\in \mathbb{R}^{n_{1}\times n_{2}\times n_{3} }   } \left \| \mathcal{X}  \right \| _{\ast }   \quad \text{s.t.} \quad \mathcal{X}_{i,j,k}= \mathcal{M}_{i,j,k},\quad \left ( i,j,k \right )\in  \Omega   
\end{equation}
where $\left \| \mathcal{X}  \right \| _{\ast }$ is the tensor nuclear norm (TNN) \cite{LJ2018}. The t-SVD decomposition and ADMM \cite{LJ2018} is one of the most popular ways to solve this problem in recent years due to thier high accuracy.

\section{\large\bf Our method of network latency estimation}

To reconstruct the whole network latencies, the entire process can be divided into two parts. Firstly, we adopts a fast and reasonable sampling strategy in Subsection A. Secondly, we estimate the rest latencies in Subsection B. Furthermore, we analyze the complexity of our algorithms in Subsection C.

\subsection{Sampling strategy}

We need to measure a few nodes before we estimate other ones in the real world. Therfore, we start with an empty tensor. By sampling, we get an incomplete tensor. Finally, we complete the incomplete tensor. Firstly, we need a reasonable and efficient sampling strategy.

Tensor leverage sampling strategy \cite{DZ2020} is an excellent approach with high accuracy because the core subroutine of this method is t-SVD that can extract the most of the information of a tensor. However, despite of high accuracy, t-SVD is not cost-efficient enough for timely feedback. Thus, we propose a new method called QRLS that is much faster than conventional leverage sampling to improve the tensor leverage sampling with faster speed.

Here are the stages of our method.

Stage 1: For the first ${t}$ time slots, namely considering a part of our tensor $\mathcal{X}(:,:,1:t)$. a random sampling strategy is adopted \cite{DZ2020}, which is the information basis for subsequent leverage sampling. $t = \left\lceil {\beta {n_3}} \right\rceil $ where $\left\lceil { \cdot  } \right\rceil $ is the ceiling brackets that rounds ${\beta {n_3}}$ up to upper integer and $\beta  \in (0,1)$. Assuming that the total number of nodes that we will measure is ${{N}}$, we will impose $\left\lceil {\frac{{\rm{N}}}{{{n_3}}}} \right\rceil $  measurement for each time slot in the first $t$ time slots.

Stage 2: For the rest ${k^{th}}$ time slot ($k \in [t + 1,{n_3}]$), we adopt a new sampling method (QRLS) that we propose. After Stage 1, we have got the sampling tensor, and then we exploit CTSVD-QR \cite{ZA2020} that is besed on tensor QR decomposition as an extension of CSVD-QR \cite{LF2019}, rather than t-SVD. The CTSVD-QR decomposition of a tensor ${\cal X}$ is as follow:

\begin{equation}
	\mathcal{X}=\mathcal{L}\ast\mathcal{D}\ast\mathcal{R}
\end{equation}

where $\mathcal{L}\in \mathbb{R}^{n_{1}\times r\times n_{3}}$,  $\mathcal{R}\in \mathbb{R}^{r\times n_{2}\times n_{3}}$ are orthogonal, and $\mathcal{D}\in \mathbb{R}^{r\times r\times n_{3}}$ is analogous to the $\mathcal{R}^{\ast}$ of Theorem \ref{T-QR}. Let ${\mathcal{D}_k}$ denotes the ${k^{th}}$ iteration of CTSVD-QR. Under constant iterations, ${\mathcal{D}_k}$ will converge to a f-diagonal tensor that is equivalent to $\mathcal{S}$ in t-SVD as Theorem \ref{T-SVD}. CTSVD-QR is an extension for the matrix decomposition CSVD-QR. In CSVD-QR, there are three basic steps as follows.

First, initialize ${L_1} = eye(m,r)$, ${D_1} = eye(r,r)$, ${R_1} = eye(r,n)$, where $eye(m,n)$ is an $m \times n$ matrix with the main diagonal element being 1 and the other positional elements being 0 in matlab function. 

Second, Perform QR decomposition on ${X_k}R_k^*$, i.e.,
\begin{equation}\label{L_k+1_qr}
	{L_{k + 1}}R' = {X_k}R_k^*
\end{equation}

${R_{k + 1}}$ is also given by QR decomposition, i.e.,
\begin{equation}
	{{R_{k + 1}}}T = A_k^*{L_{k + 1}}
\end{equation}

Eventually, we update ${D_{k + 1}}$

\begin{equation}
	{D_{k + 1}} = {T^*}
\end{equation}

CTSVD-QR can be regarded as the combination of t-SVD and CSVD. The concrete process is shown on the Algorithm \ref{algorithm1}. Note that conj(·) means the conjugate transpose operator. 

\begin{algorithm}[H]
	\caption{: Computing an Approximate T-SVD via QR Decomposition (CTSVD-QR) \cite{ZA2020}}
	\label{algorithm1}
	\begin{algorithmic}[1]
		\Require  
		$\mathcal{X}\in \mathbb{R}^{n_{1}\times n_{2}\times n_{3}}$.
		\Ensure
		CTSVD-QR components $\mathcal{L}$, $\mathcal{D}$ and $\mathcal{R}$ of $\mathcal{X}$.
		\State Compute $\hat{\mathcal{X} } =\text{fft}\left ( \mathcal{X},\left [  \right ],3   \right ) $.
		\State Compute each frontal slice of $\hat{\mathcal{L} }$, $\hat{\mathcal{D} }$ and $\hat{\mathcal{R} }$ from $\hat{\mathcal{X} }$ by
		\For{$i=1,...,\left[\frac{n_{3}+1}{2} \right]$}
		\State $\left [ \hat{\mathcal{L}}^{\left ( i \right ) }  ,\hat{\mathcal{D}}^{\left ( i \right ) },\hat{\mathcal{R}}^{\left ( i \right ) }  \right ]=\text{CSVD-QR}\left (\hat{\mathcal{X}}^{\left ( i \right ) }   \right )   $;
		\EndFor
		\For{$i=\left[\frac{n_{3}+1}{2} \right]+1,...,n_{3}$}
		\State $\hat{\mathcal{L}}^{\left ( i \right ) }=\text{conj}\left(\hat{\mathcal{L}}^{\left ( n_{3}-i+2 \right ) }\right)$;
		\State $\hat{\mathcal{D}}^{\left ( i \right ) }=\hat{\mathcal{D}}^{\left ( n_{3}-i+2 \right ) }$;
		\State $\hat{\mathcal{V}}^{\left ( i \right ) }=\text{conj}\left(\hat{\mathcal{V}}^{\left ( n_{3}-i+2 \right ) }\right)$;
		\EndFor
		\State Compute $\mathcal{L} =\text{ifft}\left ( \hat{\mathcal{L}},\left [  \right ],3   \right ) $, $\mathcal{D} =\text{ifft}\left ( \hat{\mathcal{D}},\left [  \right ],3   \right ) $, and $\mathcal{R} =\text{ifft}\left ( \hat{\mathcal{R}},\left [  \right ],3   \right ) $.
	\end{algorithmic}
\end{algorithm}

Theoretically, if we introduce CTSVD-QR into our sampling strategy, the speed will increase and the accuracy can still be maintained. According to the leverage sampling theory \cite{DZ2020}, we propose that the tensor sampling probability of measuring the $(i,j)$ node for the ${k^{th}}$ time slot can be improved by 

\begin{equation}
	{p_{i,j,k}} = \min \left\{ {{c_0}\left( {\frac{{u_i^kr}}{{{n_1}}} + \frac{{v_j^kr}}{{{n_1}}} - \frac{{u_i^kr}}{{{n_2}}}\cdot\frac{{v_j^kr}}{{{n_2}}}} \right)\log ({n_1}{n_3})\log ({n_2}{n_3}),1} \right\}
\end{equation}
where $u_i^k = \frac{n_{1}}{r}\parallel {\mathcal{L}^k}(i,1:r,:)\parallel _F^2$, $v_j^k = \frac{n_{2}}{r}\parallel {\mathcal{R}^k}(1:r,j,:)\parallel _F^2$ with the estimated rank $r$ and ${c_0}$ is a constant number for normalization. The Frobenius norm of ${i^{th}}$ horizontal slice of ${\cal L}$ and the ${j^{th}}$ lateral slice of ${\cal R}$ are respectively $u_i^k$ and $v_j^k$. 

Until $k$ is iterated to ${n_3}$, we can summarize the whole sampling strategy in Algorithm \ref{algorithm2}. Note that $\mathcal{A}$ in (\ref{Aijk}) will be constructed in Algorithm \ref{algorithm2}. We use ${\Omega _k}$ to represent the set of locations measured in the ${k^{th}}$ time slot, and ${\Omega}$ in (\ref{Aijk}) is exactly the union of sampled measurements, i.e., $\Omega  = \bigcup\limits_{k = 1}^{{n_3}} {{\Omega _k}} $.

\begin{algorithm}[H]
	\caption{: Embedding QR Decomposition into Leverage Sampling (QRLS)}
	\label{algorithm2}
	\begin{algorithmic}[1]
		\Require  
		$\mathcal{X}=\mathcal{O}\in \mathbb{R}^{n_{1}\times n_{2}\times n_{3}}$.
		\Ensure
		An incomplete tensor $\mathcal{X}$ and an updated $\mathcal{A}$. 
		\State \textbf{Stage 1: Ramdom Sampling}.
		\For{$k=1,...,t$}
		\State Randomly probe $\left\lceil {\frac{{\rm{N}}}{{{n_3}}}} \right\rceil $ nodes' latencies in the ${k^{th}}$ time slot $\mathcal{X}(:,:,k)$.
		\State Update $\mathcal{A}(:,:,k)$ : $\mathcal{A}_{ijk}{\rm{ = }}\left\{ \begin{array}{l}
			1,\quad \left( {i,j,k} \right) \in \Omega_{k} \\
			0,\quad else.
		\end{array} \right.$
		\EndFor
		\State \textbf{Stage 2: Leverage Sampling based on QR Decomposition}.
		\For{$k=t+1,...,n_3$} 
		\State $[\mathcal{L},\mathcal{D},\mathcal{R}] = $CTSVD\mbox{-}QR$(\mathcal{X}(:,:,1:k - 1))$
		\State $u_i^k = \frac{n_{1}}{r}\parallel {\mathcal{L}^k}(i,1:r,:)\parallel _F^2$
		\State  $v_j^k = \frac{n_{2}}{r}\parallel {\mathcal{R}^k}(1:r,j,:)\parallel _F^2$
		\State ${p_{i,j,k}} = \min \left\{ {{c_0}\left( {\frac{{u_i^kr}}{{{n_1}}} + \frac{{v_j^kr}}{{{n_1}}} - \frac{{u_i^kr}}{{{n_2}}}\cdot\frac{{v_j^kr}}{{{n_2}}}} \right)\log ({n_1}{n_3})\log ({n_2}{n_3}),1} \right\}$
		\State Sample $\left\lceil {\frac{{\rm{N}}}{{{n_3}}}} \right\rceil $ nodes with the order of measurement propability ${p_{i,j,k}}$ and we get $\mathcal{X}(:,:,k)$.
		\State Update the rest slices $\mathcal{A}(:,:,k)$ : $\mathcal{A}_{ijk}{\rm{ = }}\left\{ \begin{array}{l}
			1,\quad \left( {i,j,k} \right) \in \Omega_{k} \\
			0,\quad else.
		\end{array} \right.$.   
		\EndFor
	\end{algorithmic}
\end{algorithm}

\subsection{Completion for incomplete tensor}

After sampling some informative nodes, we need to estimate the latency of the rest nodes with some time slots, and the math model is tensor completion.

The general approach to Problem (\ref{1.2}) is using t-SVD to decompose $\mathcal{X}$. Due to a full rank decomposition, t-SVD need numerous iterations to solve the tensor completion problem. Iterative tensor QR decomoposition is a great way to reduce the computational time in spite of a little precision loss. 

To bypass complex tensor factorization, we use tensor ${L_{2,1}}$-norm instead of traditional nuclear norm. Here is the definition of ${L_{2,1}}$-norm of a tensor $\mathcal{X}\in\mathbb{R}^{n_{1}\times n_{2}\times n_{3}}$ :

\begin{equation}\label{L2,1-norm of tenosr}
	{\left\| {\cal X} \right\|_{2,1}} = \mathop \sum \nolimits_{j = 1}^{{n_2}} \sqrt {\mathop \sum \nolimits_{i = 1}^{{n_1}} \mathop \sum \nolimits_{k = 1}^{{n_3}} {\cal X}_{ijk}^2}  = \mathop \sum \nolimits_{j = 1}^{{n_2}} {\left\| {\mathcal{X}(:,j,:)} \right\|_F}
\end{equation}

\begin{theorem}\label{L21}\cite{ZA2020}
	Assuming that there is a tensor $\mathcal{X}\in \mathbb{R}^{n_{1}\times n_{2}\times n_{3}}$, imposing CTSVD-QR on $\mathcal{X}$, namely $\mathcal{X}=\mathcal{L}\ast \mathcal{D}\ast\mathcal{R}$, we have $\left \| \mathcal{D}  \right \| _{\ast}\le \left \|\mathcal{D}   \right \| _{2,1}$.
\end{theorem}

Theorem \ref{L21} demonstrates that for tensor $\mathcal{D}$, the nuclear norm is the lower bound of the ${L_{2,1}}$-norm. Thus, Problem (\ref{1.2}) can be transformed as follows:

\begin{equation}\label{16}
	\min_{\mathcal{D} } \left \| \mathcal{D}  \right \| _{2,1}   \quad \text{s.t.} \begin{cases}
		\mathcal{X}=\mathcal{L}\ast \mathcal{D}\ast \mathcal{R}    \\
		\mathcal{X}_{i,j,k}= \mathcal{M}_{i,j,k}
	\end{cases}.
\end{equation}

The objective function in Problem (\ref{16}) is convex \cite{ZA2020}, so we can use ADMM to solve the problem. The augmented Lagrange function of Problem (\ref{16}) is  

\begin{equation}
	\text{Lag}\left ( \mathcal{L},\mathcal{D},\mathcal{R},\mathcal{Y},\mu    \right )   =\left \| \mathcal{D}  \right \| _{2,1}+\left \langle \mathcal{Y},\mathcal{X} - \mathcal{L}\ast\mathcal{D}\ast\mathcal{R}    \right \rangle  +\frac{\mu }{2}\left \|\mathcal{X} - \mathcal{L}\ast\mathcal{D}\ast\mathcal{R}  \right \|  _{F}^{2} ,
\end{equation}
where $\mu  > 0$, $\mathcal{Y}\in \mathbb{R}^{n_{1}\times n_{2}\times n_{3}}$, and $\langle \mathcal{X},\mathcal{Y}\rangle $ is the inner product of $\mathcal{X}$ and $\mathcal{Y}$, i.e.,
\begin{equation}
	\langle \mathcal{X},\mathcal{Y}\rangle  = \sum\limits_{i = 1}^{{n_3}} {\langle {\mathcal{X}^{(i)}},{\mathcal{Y}^{(i)}}\rangle } 
\end{equation}

Let ${\mathcal{X} _k}$ denotes the ${k^{th}}$ iteration result in ADMM. We optimize the Lagrange function in the following three steps.

In the first step, we update ${\mathcal{L} _{k+1}}$ and ${\mathcal{R} _{k+1}}$ by solving the following optimization problem:
\begin{equation}\label{3}
	\min_{\mathcal{L},\mathcal{R} } \left \| \left ( \mathcal{X}_{k}+\frac{\mathcal{Y}_{k} }{\mu _{k}} \right ) -\mathcal{L}\ast \mathcal{D}_{k}\ast \mathcal{R} \right \| ^{2 } _{F}
\end{equation}

The optimization Problem in (\ref{3}) is convex to each variable, so we can update the variables one by one. We use ADMM to initial $\mathcal{L}_k$ and $\mathcal{R}_k$ as the first iteration in CTSVD-QR. Due to converging within a few iterations, ${\mathcal{L} _{k+1}}$ and ${\mathcal{R} _{k+1}}$ can be given by tensor QR decomposition. We perform tensor QR decomposition on $\mathcal{L}_{k + 1}$ and $\mathcal{R}_{k + 1}$, i.e.,
\begin{equation}
	{\mathcal{L}_{k + 1}}\mathcal{R'} = {\rm{t\mbox{-}QR}}\left( {({\mathcal{X}_k} + \frac{{{\mathcal{Y}_k}}}{{{\mu _k}}}) * \mathcal{R}_k^ * } \right)
\end{equation}
\begin{equation}\label{R''}
	{\mathcal{R}_{k + 1}}\mathcal{R''} = {\rm{t\mbox{-}QR}}\left( {{{({\mathcal{X}_k} + \frac{{{\mathcal{Y}_k}}}{{{\mu _k}}})}^ * } * {\mathcal{L}_{k + 1}}} \right)
\end{equation}
where $t\mbox{-}{\rm{QR}}$ denotes the tensor QR decomposition operator.

In the second step, ${\mathcal{D}_{k + 1}}$ and ${\mathcal{X}_{k + 1}}$ are updated with ${\mathcal{X}_k}$, ${\mathcal{Y}_k}$, ${L_{k + 1}}$ and ${R_{k + 1}}$ being fixed. We attempt to decompose ${({\mathcal{X}_k} + \frac{{{\mathcal{Y}_k}}}{{{\mu _k}}})}$ in one iteration of CTSVD-QR, i.e.,
\begin{equation}\label{init}
	{\mathcal{X}_k} + \frac{{{\mathcal{Y}_k}}}{{{\mu _k}}} = {\mathcal{L}_{k + 1}}*{\mathcal{D_T}}*{\mathcal{R}_{k + 1}}
\end{equation}
where $\mathcal{D}_T\in \mathbb{R}^{r\times r\times n_{3}}$. Then, we consider a ${L_{2,1}}$-norm minimization problem to update ${D_{k + 1}}$, i.e.,
\begin{equation}\label{min2}
	\mathcal{D} _{k+1}=\text{arg}\min_{\mathcal{D} }  \frac{1}{\mu _{k}}\left \| \mathcal{D}  \right \|  _{2,1} +\frac{1}{2}\left \| \mathcal{D} -\mathcal{L}^{\ast }_{k+1} \ast \mathcal{X}_{c} \ast\mathcal{R}^{\ast }_{k+1}  \right \|^{2}_{F}  .
\end{equation}

According to (\ref{init}), we can get  $\mathcal{D_T}$:
\begin{equation}\label{DT}
	{\mathcal{D_T}} = \mathcal{L}_{k + 1}^**({\mathcal{X}_k} + \frac{{{\mathcal{Y}_k}}}{{{\mu _k}}})*\mathcal{R}_{k + 1}^*
\end{equation}

From (\ref{R''}) and (\ref{DT}), we can know that $R''=\mathcal{D_T}$.

Suppose that the tensor $\mathcal{D}$ can be decomposed as follows:
\begin{equation}\label{def1}
	\mathcal{D} = \sum\limits_{k = 1}^r {{\mathcal{D}^k}}, 
\end{equation}
\begin{equation}\label{def2}
	{\cal D}_{ijk} = {\delta _{j,k}}\cdot{\cal D}_{ijk} = \left\{ {\begin{array}{*{20}{c}}
			{{\cal D}_{ijk},\quad j = k}\\
			{0,\quad j \ne k}.
	\end{array}} \right.
\end{equation}

From (\ref{L2,1-norm of tenosr}), (\ref{def1}) and (\ref{def2}), we can simplify the Problem(\ref{min2}) as a Frobenius norm minimization problem:
\begin{equation}\label{problem D^j_k+1}
	\hat{\mathcal{D}}^{j} _{k+1}=\text{arg}\min_{\hat{\mathcal{D}}^{j} }  \frac{1}{\mu _{k}}\left \| \hat{\mathcal{D}}^{j}  \right \|  _{F} +\frac{1}{2}\left \| \hat{\mathcal{D}}^{j} -\hat{\mathcal{D_T}}^{j}  \right \|^{2}_{F},\mathcal{\widehat{D}} = {\bf{fft}}(\mathcal{D},3)
\end{equation}

The ${l^{th}}$ frontal slice of $\hat{\mathcal{D}}^{j} _{k+1}$ can be represented as follow \cite{ZA2020}:

\begin{equation}\label{D_k+1}
	\mathcal{D}{_{k + 1}^j}^{(l)} = {\bf{ifft}}\left( {\max \left\{ {{{\left\| {{{\widehat {\mathcal{D_T}^j}}^{(l)}}} \right\|}_F} - \frac{1}{\mu },0} \right\} \cdot \frac{{{{\widehat {\mathcal{D_T}^j}}^{(l)}}}}{{{{\left\| {{{\widehat {\mathcal{D_T}^j}}^{(l)}}} \right\|}_F}}},3} \right),l = 1, \cdots ,{n_3}
\end{equation}

The iteration speed of (\ref{D_k+1}) will be much faster than traditional tensor decomposition. And then we update ${\mathcal{X}_{k + 1}}$ as follow:
\begin{equation}
	{{\cal X}_{k + 1}} = {{\cal L}_{k + 1}} * {{\cal D}_{k + 1}} * {{\cal R}_{k + 1}} - \left( {{{\cal L}_{k + 1}} * {{\cal D}_{k + 1}} * {{\cal R}_{k + 1}}} \right) \circ \mathcal{A} + M \circ \mathcal{A}
\end{equation}
where $\mathcal{A}$ is mentioned in (\ref{Aijk}). 

In the final step, we update $\mathcal{Y}_{k+1}$ and $\mathcal{\mu}_{k+1}$ with ADMM method:
\begin{equation}
	\mathcal{Y}_{k+1}=\mathcal{Y}_{k}+\mu_{k}\left(\mathcal{X}_{k+1}-\mathcal{L}_{k+1}\ast\mathcal{D}_{k+1}\ast\mathcal{R}_{k+1}\right)
\end{equation}
\begin{equation}
	\mu_{k+1}=\rho \mu_{k},\quad \rho\ge 1
\end{equation}

The whole tensor completion process above can be regarded as a modified ADMM method. Combining the QRLS sampling mathod that we propose, we can get the global algorithm for network latency estimation problem called LNLS-TQR in Algorithm \ref{algorithm3}. 

\begin{algorithm}[H]
	\caption{: Tensor ${L_{2,1}}$-Norm minimization and Leverage Sampling based on Tensor QR Decomposition (LNLS-TQR)}
	\label{algorithm3}
	\begin{algorithmic}[1]
		\Require  
		$\mathcal{X}=\mathcal{O}\in \mathbb{R}^{n_{1}\times n_{2}\times n_{3}}$.
		\Ensure
		The recovered tensor $\mathcal{X}$.	
		\State Initialize: The tubal rank $r>0$. $k=0$, $c_{0}>0$, $l>0$, $\mu>0$, $\rho>0$. The positive tolerance $\epsilon >0$,  $\mathcal{Y}=\mathcal{O}\in \mathbb{R}^{n_{1}\times n_{2}\times n_{3}}$, $\mathcal{L}_{1}=\mathcal{I}\in \mathbb{R}^{n_{1}\times r\times n_{3}}$, $\mathcal{D}_{1}=\mathcal{I}\in \mathbb{R}^{r\times r\times n_{3}}$, $\mathcal{R}_{1}=\mathcal{I}\in \mathbb{R}^{r\times n_{2}\times n_{3}}$, $\mathcal{A}=\mathcal{O}\in \mathbb{R}^{n_{1}\times n_{2}\times n_{3}}$.
		\State $[\mathcal{X}_1, \mathcal{A} ]=$QRLS$(\mathcal{X})$.
		
		\While{$\left \|\mathcal{L}_{k}\ast\mathcal{D}_{k}\ast\mathcal{R}_{k}-\mathcal{X}_{k}\right \| ^{2}_{F}\ge \epsilon \quad \text{and}\quad k>l $}
		\State$[{L_{k + 1}}, \sim ]  = \rm{t\mbox{-}QR}\left( {({\mathcal{X}_k} + \frac{{{\mathcal{Y}_k}}}{{{\mu _k}}}) * \mathcal{R}_k^ * } \right)$
		\State $\left [ \mathcal{R}_{k+1},\mathcal{D_T}^{\ast}    \right ] =\rm{t\mbox{-}QR}\left(\left(\mathcal{X}_{k}+\frac{\mathcal{Y}_{k} }{\mu _{k}}\right)^{\ast}\ast \mathcal{L}_{k+1} \right)$
		\For{$t=1,...,n_{3}$}
		\For{$j=1,...,r$}
		\State $\mathcal{D}{_{k + 1}^j}^{(l)} = {\bf{ifft}}\left( {\max \left\{ {{{\left\| {{{\widehat {\mathcal{D_T}^j}}^{(l)}}} \right\|}_F} - \frac{1}{\mu },0} \right\} \cdot \frac{{{{\widehat {\mathcal{D_T}^j}}^{(l)}}}}{{{{\left\| {{{\widehat {\mathcal{D_T}^j}}^{(l)}}} \right\|}_F}}},3} \right),\quad l = 1, \cdots ,{n_3}$
		\EndFor
		\EndFor
		\State $\mathcal{X}_{k+1}=\mathcal{L}_{k+1}\ast\mathcal{D}_{k+1}\ast\mathcal{R}_{k+1}-\left( {{{\cal L}_{k + 1}} * {{\cal D}_{k + 1}} * {{\cal R}_{k + 1}}} \right) \circ \mathcal{A}+\mathcal{X}_{1}$;
		\State $\mathcal{Y}_{k+1}=\mathcal{Y}_{k}+\mu_{k}\left(\mathcal{X}_{k+1}-\mathcal{L}_{k+1}\ast\mathcal{D}_{k+1}\ast\mathcal{R}_{k+1}\right)$;
		\State $\mu_{k+1}=\rho \mu_{k}$;		
		\EndWhile
		\State \Return $\mathcal{X}=\mathcal{X}_{k}$.
	\end{algorithmic}
\end{algorithm}

\subsection{Analysis of computational complexity}

In Algorithm 2, the main complexity of QRLS mainly concentrate on Step 8 rather than computing the leverage scores, which costs $O\left( {({n_1} + {n_2}){r^2}{n_3}} \right)$ \cite{ZA2020}. In the application of network estimation, the complexity will be $O\left( {2n{r^2}{n_3}} \right)$ because $n=n_{1}=n_{2}$. The complexity of tensor leverage sampling is $O\left( {({n_3} - 1)(2nr - {r^2}){{\log }^2}(n{n_3})} \right) \approx O\left( {2rn{n_3}\log (n{n_3})} \right)$ \cite{DZ2020}, which is not the main cost of QRLS. Thus, the per-iteration complexity of QRLS in network latency estimation is $O\left( {2n{r^2}{n_3} + ({n_3} - 1)(2nr - {r^2}){{\log }^2}(n{n_3})} \right)$.

In Algorithm 3, the main cost is in step 4, step 5, step 11 and step 12. Step 4 costs $O\left( {{n_1}{n_2}r{n_3} + 2{n_2}{r^2}{n_3}} \right)$ and step 5 costs $O\left( {{n_1}{n_2}r{n_3} + 2{n_1}{r^2}{n_3}} \right)$, namely $O\left( {{n^2}r{n_3} + 2n{r^2}{n_3}} \right)$ in our application. Step 11 and step 12 costs $O\left( {{n_1}{n_2}r{n_3} + {n_1}{r^2}{n_3}} \right)$, namely $O\left( {{n^2}r{n_3} + n{r^2}{n_3}} \right)$ in our application.

%
\section{Experiments}

To demonstrate the feasibility of our approach, several comparative experiments including synthetic and real datasets, were performed. The results were compared with the running time and the relative square error (RSE), i.e., ${\rm{RSE}} = \sqrt {\frac{{\left\| {{\cal Y} - {\cal X}} \right\|_F^2}}{{\left\| {\cal X} \right\|_F^2}}} $, where $\mathcal{X}, \mathcal{Y}\in\mathbb{R}^{n_{1}\times n_{2}\times n_{3}} $ are the completed tensor and the real tensor respectively.

We perform our method on MATLAB 2021b platform equipped with an Intel (R) Core (TM) i7-11800H CPU and 32.0 GB of RAM.

\subsection{Experiments with a synthetic dataset}

Since the tensor model of network latency estimation has been proved to have the property of approximately low rank \cite{DZ2020}, we can use tensor completion method to solve this problem. Besides, the key subroutine of our method CTSVD-QR has also been proved convergent \cite{ZA2020}.

For the original tensor $\mathcal{X}\in\mathbb{R}^{n_{1}\times n_{2}\times n_{3}} $, we choose ${n_1} = {n_2} = 50$, and ${n_3} = 10$. We normalized the data before complete the tensor. In our method, we choose $\mu {\rm{ = }}0.01$, $\beta {\rm{ = }}0.10$  and $\rho {\rm{ = }}1.50$. In addition, we test whether the tensor completion results will be disturbed greatly under Gaussian noise $\mathcal{N}\in\mathbb{R}^{n_{1}\times n_{2}\times n_{3}} $ where $\mathcal{N} \sim \mathcal{N}(\mu ,\sigma )$ with $\mu  = 0$, and $\sigma  = 0.01$. 

\begin{figure}
	\centering
	\includegraphics[width=0.8\linewidth]{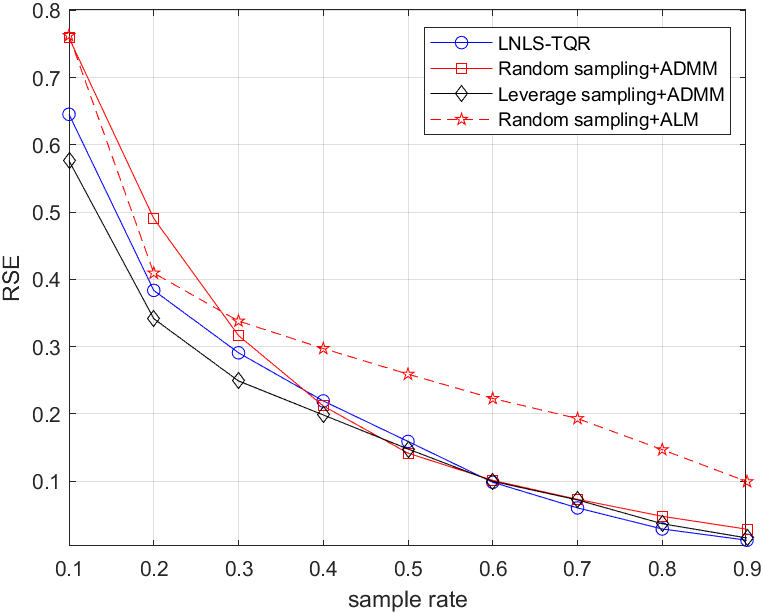}
	\caption{Accuracy for the synthetic dataset without noise}
	\label{fig:comparisionsyneticdatawithoutnoise}
\end{figure}

The speed of the algorithm is represented by the interval of one iteration of an algorithm, and the accuracy of the algorithm is represented by RSE with the changeable sampling rate with tubal-rank $r=5$. Table \ref{Running time} demonstrates the running time of one iteration of the dataset through recent four algorithms equipped with speed and precision. Figure \ref{fig:comparisionsyneticdatawithoutnoise} depicts the average accuracy of the four algorithms for numerous iterations. 

\begin{table}[H]
	\caption{Running time and RSE value for the synthetic dataset  in average.}
	\label{Running time}
	\begin{center} 
		\begin{tabular}{lcc}
			\toprule
			Algorithm & Running Time (seconds)   &  RSE value\\
			\midrule
			LNLS-TQR            & \pmb{0.03334}   & 0.2107   \\
			Random sampling+ADMM                & 0.2716  &  0.2409   \\
			Leverage sampling+ADMM            & 0.2923    & \pmb{0.1928}   \\
			Random sampling+ALM          & 4.5411      & 0.3030  \\
			\bottomrule
		\end{tabular}
	\end{center}
\end{table}


\begin{figure}
	\centering
	\includegraphics[width=0.8\linewidth]{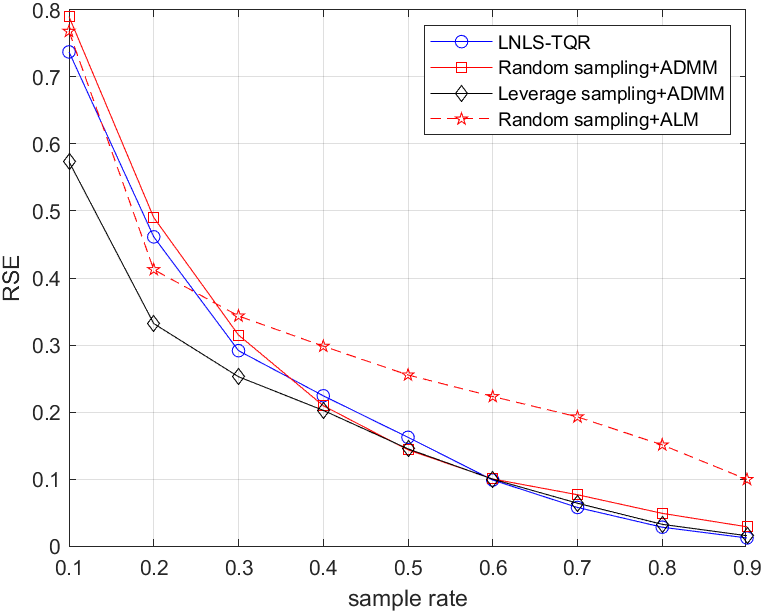}
	\caption{Accuracy for the synthetic dataset with noise}
	\label{fig:comparisionsyneticdata3noise}
\end{figure}

It is clear that our method is obviously much faster than others while the precision of it is also satisfactory. Our method performs great at high sampling rates and performs acceptable at low sampling rates. The results is in line with expectations because the speed of tensor QR decomposition(t-QR) is $\mathit{O}\left(2n_{1}n_{2}^{2}n_{3} \right)$ while that of t-SVD is $\mathit{O}\left(2n_{1}n_{2}^{2}n_{3} \right)$, which means t-QR will fater than t-SVD for nearly 10 times. In addition, t-QR is used to approximate t-SVD. Thus, t-QR will not be more accurate than t-SVD but the effect of approximation is satisfactory. Figure \ref{fig:comparisionsyneticdata3noise} indicates that our method suffers very little from the noise.

\subsection{Experiments with a real dataset}

Here, our method will be evaluated by the real Seattle dataset \cite{DZ2020,CI2009}. The Seattle latency dataset includes the round-trip times (RTTs) between 99 personal devices over 3 hours in Seattle \cite{CI2009}. There are 688 latency matrices in this dataset that become a third order tensor ${\cal X} \in {\mathbb{R}^{99 \times 99 \times 688}}$ and the low-rank property has been fully proved \cite{DZ2020}. We randomly choose 10 sequential frontal slices of the tensor in the Seattle dataset, namely 10 groups of sequential time frames. 

\begin{figure}
	\centering
	\includegraphics[width=0.8\linewidth]{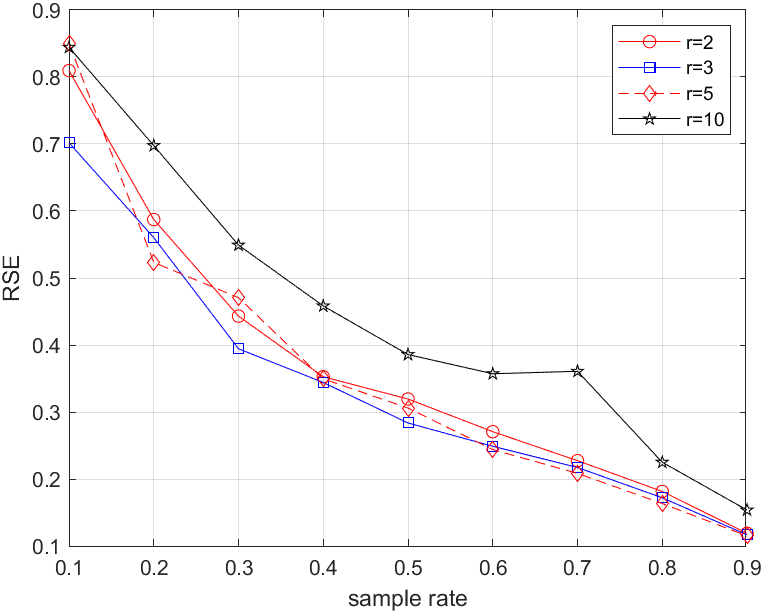}
	\caption{The relationship between estimated rank and RSE}
	\label{fig:rank2}
\end{figure}

Since getting the totally precise tubal-rank of a tensor is a NP-hard problem and the time slots will not be too much in the estimation of network latency, we estimate the tubal-rank by experiments from 1 to 10 in Figure \ref{fig:rank2}. We can see that our algorithm performs well when the rank is 1 to 5 and we can judge that the approximate tubal-rank of this dataset is in this range. $r=3$ is the best case, but in fact, we may not be able to find the optimal estimated rank in all cases, so in the following experiment, we choose $r=2$ with medium performance as our estimated rank.

\begin{figure}
	\centering
	\includegraphics[width=0.8\linewidth]{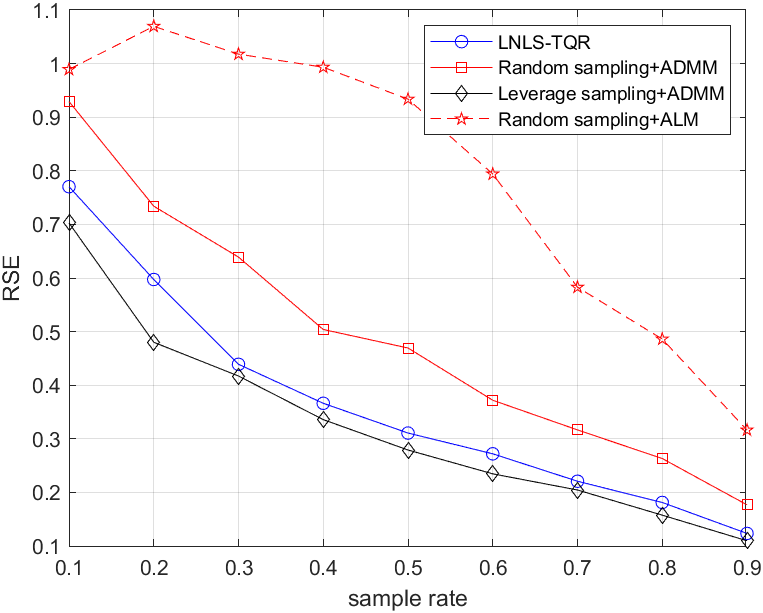}
	\caption{Completion error for the Seattle dataset without noise}
	\label{Completion error for the Seattle dataset without noise}
\end{figure}

\begin{figure}
	\centering
	\includegraphics[width=0.8\linewidth]{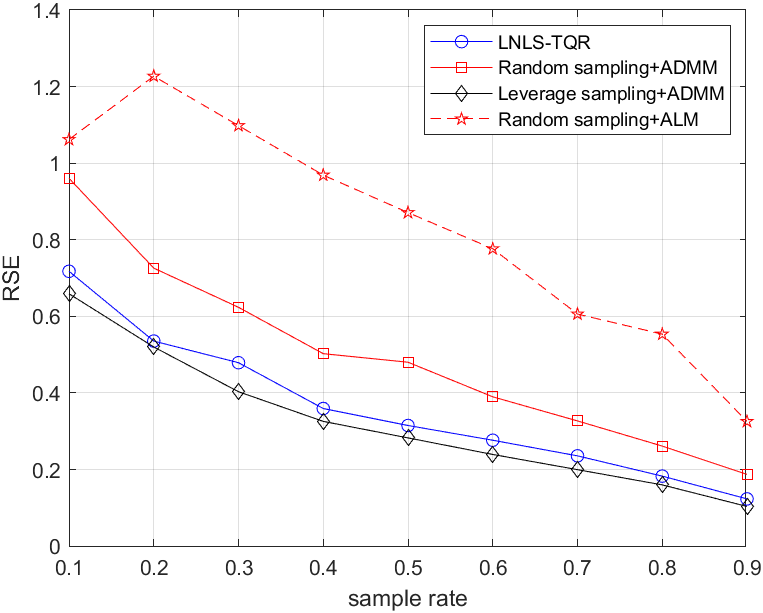}
	\caption{Completion error for the Seattle dataset with noise}
	\label{fig:comparisionseattledatawithnoise}
\end{figure}

For the sake of the optimal effect of the algorithms, we still choose $\mu {\rm{ = }}0.01$, $\beta {\rm{ = }}0.10$ and $\rho {\rm{ = }}1.50$. The comparison result of recent algorithms in terms of speed is shown in Table \ref{Running time2} and the comparison conclusion of recent algorithms in terms of accuracy is shown in Figure \ref{Completion error for the Seattle dataset without noise} and Figure \ref{fig:comparisionseattledatawithnoise}. We can judge that our algorithm is obviously faster than others with a satisfied precision. At the same time, our algorithm also has high accuracy, especially when the sampling rate is greater than or equal to 0.4. The precision at low sampling rate is also satisfactory. Figure \ref{fig:comparisionseattledatawithnoise} also illustrate that our method has a strong resistance to noise.  

\begin{table}[H]
	\caption{Running time and RSE value for the real dataset  in average.}
	\label{Running time2}
	\begin{center} 
		\begin{tabular}{lcc}
			\toprule
			Algorithm & Running Time (seconds)   &  RSE value\\
			\midrule
			LNLS-TQR            & \pmb{0.1011}   & 0.3647   \\
			Random sampling+ADMM                & 0.9798  &  0.4894   \\
			Leverage sampling+ADMM            & 1.041    & \pmb{0.3247}   \\
			Random sampling+ALM          & 7.244      & 0.7980  \\
			\bottomrule
		\end{tabular}
	\end{center}
\end{table}

\section{Conclusion}
In this papper, we propose a new method for network latency estimation. This method includes the fast leverage sampling strategy and tensor QR decomposition. The network latency estimation can be abstracted into a tensor completion problem. The purpose of the fast leverage sampling is to collect more informative nodes efficiently, and tensor QR decomposition is to approximate t-SVD that is a critical subroutine of tensor decomposition with high precision, and then the tensor completion problem can be solved better. The advantage of our method is that it has several times the speed up and little accuracy loss compared to t-SVD and tensor ADMM. In addition, we use synthetic and real-world datasets to evaluate our method. The experiment witnesses the fast speed and high accuracy of our method.

%
%
%
%
%
%
%

%
%
%
%
%

\end{document}